\documentclass[%
 reprint,
 amsmath,amssymb,mathrsfs
 aps,
]{revtex4-2}

\usepackage{graphicx}
\usepackage{dcolumn}
\usepackage{bm}
\usepackage{comment}

\begin{document}

\preprint{APS/123-QED}

\title{Beam-Beam Interaction in a Dielectric Laser Accelerator Electron-Positron Collider}

\author{R. Joel England}
\email{england@slac.stanford.edu}
\affiliation{SLAC National Accelerator Laboratory, 2575 Sand Hill Rd, Menlo Park, CA 94025-7015 USA}

\author{Levi Sch\"{a}chter}
\email{levi@ee.technion.ac.il}
\affiliation{Technion - Israel Institute of Technology, Haifa 32000, Israel}

\begin{abstract}

We examine through numerical calculation the collision of counter-propagating  trains of optically spaced electron/positron microbunches in a 1 TeV collider scenario for a dielectric laser accelerator (DLA). A time-dependent envelope equation is derived for arbitrary number of bunches in the classical limit, with inclusion of the radiation reaction force (RRF).  Example parameters are examined based on a constrained luminosity relation that takes into account the bunch charge for optimal efficiency, material damage limits, and power constraints. We find that for initially identical counter-propagating Gaussian bunch trains the periodic temporal structure leads to a peak in luminosity with number of bunches. For longer bunch trains, the enhancement then decreases inversely with number of bunches. The corresponding fractional energy loss of the beam is found to be of order 1.75\%, which is reduced to 0.35\% when the nonlinear radial dependence of the transverse force is included, with an average beamstrahlung parameter of 0.075, an important result considering that beamstrahlung losses are a critical concern for future TeV colliders. 
\end{abstract}

\maketitle


\section{\label{sec:intro}Introduction}

A future electron-positron linear collider is one of the primary proposed tools for experimental investigation of high energy physics beyond the limits of the Large Hadron Collider (LHC). At present, the two main proposals for a next generation electron/positron collider are the International Linear Collider (ILC) and the Compact Linear Collider at CERN (CLIC). Both approaches are based on well established radio frequency (RF) technology, with the first (ILC) relying on superconducting microwave technology and the second (CLIC) on a normal-conducting two-beam acceleration scheme \cite{ILC2013,CLIC2012}. However, heat-induced quenching of superconducting cavities and surface breakdown for metallic RF structures limit the achievable gradients in these approaches to of order 30 and 100 MeV/m respectively. Consequently, many kilometers of real estate (and tens of billions of U.S. dollars) are required to reach TeV scale energies. The enormous cost and footprint requirements for future colliders has prompted interest in the development of new acceleration techniques with the potential for higher gradients and lower cost. 

Of these alternative schemes, the most heavily explored to date is the plasma wakefield accelerator, which uses either a high-peak-power laser pulse (laser wakefield accelerator or LWFA) or a high-density particle beam (plasma wakefield accelerator or PWFA) to generate a longitudinal plasma wave that can accelerate one or more trailing bunches \cite{tajima:1979,rosenz:1987}. Plasma-based accelerators have shown enormous progress, with demonstrated acceleration gradients in excess of 40GeV/m \cite{esarey:2009,blumenfeld:2007,gonsalves:2019}. Consequently, the length of such a machine could potentially be on the order of a hundred meters. However, efficiency of the drive laser and/or plasma instabilities at the required repetition rates (of order 10 kHz) pose significant technical challenges \cite{muggli:2012,Huntington:2011,Schroeder:2010}. 
\begin{table}[b]
\caption{\label{tab:table1}%
Parameters leading to ILC \cite{ILC2013} and DLA \cite{hanuka:regimes:2018,hanuka:single:2018,hanuka:multi:2018} geometrical (unenhanced) luminosity rate of $\bar{\mathcal{L}}_\text{ref}$ = $2 \times 10^{34}$ cm$^{-2}s^{-1}$ at 1 TeV center of mass energy.
}
\begin{ruledtabular}
\begin{tabular}{lcc}
\textrm{Parameter}&
\textrm{ILC}&
\textrm{DLA}\cr
\colrule
Bunch length (mm) & 0.3 & 0.0004 \\
Beam Size X (nm) & 474 & 0.76  \\
Beam Size Y (nm) & 5.9 & 0.76 \\
Bunch population $N$ & $2\times10^{10}$ &  $2\times10^{6}$ \\
Bunches (M) per crossing & 1 & 100\\
Bunches per train & 1312 & 100\\
Bunch train repetition rate (Hz) & 5 & $168\times10^{6}$\\
\end{tabular}
\end{ruledtabular}
\end{table}

We consider here a different approach, based on dielectric laser acceleration (DLA). The technique utilizes a slow wave hollow-core dielectric micro-structure for acceleration, driven by an infrared laser pulse. In recent experiments, measured acceleration gradients up to 850 MeV/m have been demonstrated with sustained axial fields of 1.8 GV/m \cite{peralta:2013,wootton_demonstration_2016,cesar_nonlinear_2018}. Compatible laser-driven focusing methods for transverse particle confinement over long distances have also been developed \cite{naranjo_stable_2012,niedermayer:focusing:2018,black:2019}. With the increase in efficiency of wall-plug to laser light over the past two decades, the projected machine efficiency for a DLA collider is comparable with that of ILC or CLIC \cite{england:rmp2014}. A conceptual roadmap for a future multi-TeV DLA collider is considered among various other advanced schemes in Ref. \cite{ALEGRO:2019}, with outlines of the various technical constraints pertaining to each approach. While the stringent demands of a HEP collider pose significant challenges for any accelerator technology, we note that the scope of the present work is not to address all of the technical issues of a DLA collider, but rather to study one particular concern, namely the multi-bunch luminosity and energy loss at the interaction point (IP). 

One of the main concerns associated with a DLA based collider is whether it can deliver the required luminosity, since the beam parameters are very different than for either conventional RF or plasma based schemes \cite{hanuka:regimes:2018,hanuka:single:2018,hanuka:multi:2018}. For the DLA case, beam-loading and efficiency considerations limit the macrobunch population to order of $10^6$ particles whereas space-charge limitations require splitting the macro-bunch into a train of microbunches with separation determined by the laser wavelength (here $\lambda$ = 2 $\mu$m). These microbunches naturally have sub-fs bunch duration, as has recently been demonstrated in net acceleration experiments \cite{black:Atto:2019,schoenenberger:Atto:2019}. Consequently, the laser repetition rate must be of order 10 to 100 MHz in order to deliver the required average current. This requirement is readily met by current solid state fiber lasers, which have compact footprint (e.g. rack-mounted), $\mu$J pulse energies, and multi-MHz repetition rates, in contrast to the high peak power lasers used for plasma acceleration \cite{Schroeder:2010}. In Table \ref{tab:table1}, we present a typical set of parameters for the ILC \cite{ILC2013} and DLA cases such that in both cases the bare luminosity is the same. The superscripted DLA$^1$ and DLA$^2$ denote the two test cases (Case 1 and Case 2) discussed in Section \ref{sec:params}. 

In the DLA scenario, the entire train of $M$ microbunches cross together at the IP, producing a total of $M^2$ crossings per laser pulse, whereas in RF or plasma-based schemes, the bunches in each train are sufficiently far apart that only one pair of bunches interact at a given time. This difference in bunch format has significant implications for the beam-loaded efficiency, luminosity, and energy loss dynamics. For a given repetition rate and number of particles the primary way to enhance the luminosity is to reduce the transverse beam size at the IP. This happens naturally as electrons and positrons collide, producing a pinching effect due to the lensing of each beam's charge distribution on the other. 

It is our goal in this study to investigate this process for colliding electron-positron bunch trains with DLA like parameters. Both electrons and positrons are assumed to be equally affected (strong-strong interaction) as each microbunch experiences the quasi-periodic force experienced by the counter-propagating species. As each microbunch is deflected by the counter-propagating train, radiation (beamstrahlung) is emitted. We note that our analytical formulation is agnostic as to which of the two beams contains electron vs. positrons. Since the colliding beams are identical, the resulting beam envelope analysis applies to both beams, given that the relative charge between them has the product $-e^2 N^2$. Our analysis considers the impact of the bunch train structure on the emitted radiation, but we ignore $e^+$ $e^-$ pair production.  

In the following section, we derive the relation for the luminosity as a function of number of microbunches $M$ corresponding to optimal beam-loading efficiency. We then subject this relation in Section \ref{sec:params} to constraints upon the material damage limits, power consumption, and bunch charge to obtain two sets of example collider parameters for further analysis. We then derive the multi-bunch equations of motion and energy loss relations in Sections \ref{sec:eqsofmotion}, \ref{sec:RRFeqns}, and \ref{sec:upsilon}. In Section \ref{sec:results} we present numerical solutions for the considered beam parameters up to $M = 150$, followed by analysis and concluding remarks in Section \ref{sec:conclusions}.

\section{Microbunch Efficiency}
\label{sec:efficiency}
For a multi-bunch beam with a constant \textit{unloaded} gradient $G_0$, the corresponding \textit{loaded} gradient $G_L$ at a given bunch position will vary from bunch to bunch due to the wakes from prior bunches, resulting in energy spread between the bunches. In order to mitigate this effect and produce a constant loaded gradient, the laser pulse may be tapered so as to counteract the bunch wakes and level out the loaded gradient. As discussed in \cite{hanuka:multi:2018}, the axial longitudinal field in this scenario has the form
\begin{equation}
    E_z(z,t) = \mathcal{E}(z,t) \cos (\omega t)
\end{equation}
where the laser field amplitude function $\mathcal{E}(z,t)$ is a flat pulse of amplitude $\mathcal{E}_0$ over a duration $\tau_D$ followed by a step of duration $\tau_B$ and magnitude $(1+\delta)\mathcal{E}_0$. The wake correction factor $\delta \equiv 2 k q L / (\mathcal{E}_0 \lambda)$ is derived in Ref. \cite{hanuka:multi:2018}. Here $q$ is the micro-bunch charge, $\lambda$ is the laser wavelength and microbunch spacing, and $L$ is the length of a single acceleration stage. We additionally introduce the fundamental mode loss factor $k$ and the Cherenkov wake loss factor $h$, given by \cite{siemann:2004,na:2005}
\begin{equation}
    k = \frac{1}{4} \frac{c Z_C}{\lambda^2} \frac{\beta_g}{1-\beta_g} \ \ ; \ \ h = \frac{c Z_H}{\lambda^2}
\end{equation}
where $Z_C$ is the characteristic impedance of the fundamental mode, $Z_H$ is the Cherenkov wake impedance, and $\beta_g = v_g /c$ is the group velocity normalized to light speed. The interaction time is limited by the number of bunches $M$ and laser pulse duration $\tau_P = \tau_B + \tau_D$ where 
\begin{equation}
    \tau_B = (M-1)\lambda/c \ \ ; \ \ \tau_D = \frac{1-\beta_g}{\beta_g} \frac{L}{c} \ .
\end{equation}

The corresponding unloaded and loaded gradients seen by the $i$'th bunch are then given by
\begin{equation}
    G_0(i) = \mathcal{E}_0+(i-1)2 k q \ ; \ G_L = \mathcal{E}_0- h q
\end{equation}
where the extra factor of $(i-1)2 k q$ is the wake correction to flatten the loaded gradient. The unloaded gradient averaged over all $M$ bunches is then $\langle G_0 \rangle = \mathcal{E}_0+k q (M-1)$. A detailed treatment is discussed in \cite{hanuka:multi:2018}. We will not repeat the derivation there but merely note that the corresponding $M$-dependent beam-loaded energy transfer efficiency is given by
\begin{equation}
\label{eqn:efficiency}
    \eta(q,M) = \frac{4 M k q (\mathcal{E}_0 - h q)}{\mathcal{E}_0^2 [1+\frac{\beta_g}{1-\beta_g} (M-1) \frac{\lambda}{L}(1+\frac{2kq}{\mathcal{E}_0} \frac{L}{\lambda})^2]}
\end{equation}
We can obtain from this the optimal charge $q_\text{opt}$ corresponding to maximum efficiency $\eta_\text{opt}$ by solving $\partial\eta / \partial q = 0$ for $q$, which gives
\begin{equation}
\label{eqn:optcharge}
    q_\text{opt} = \frac{\mathcal{E}_0}{h (1+\chi)} \ , 
\end{equation}
\begin{equation}
\label{eqn:optefficiency}
    \eta_\text{opt} = \frac{4 M}{1+\chi} \frac{\eta_\text{sc} (\frac{\chi}{\chi+1})}{1+\kappa \xi (1+\frac{2 \eta_\text{sc}}{\xi (1+\chi)})^2}
\end{equation}
where $\eta_\text{sc} \equiv k/h$ is the single-bunch efficiency, $\xi \equiv \lambda/L$ and for the sake of compactness we define the dimensionless parameters
\begin{equation}
    \kappa \equiv \frac{\beta_g}{1-\beta_g} (M-1) \ ; \ \chi \equiv \sqrt{\frac{\xi + \kappa (2 \eta_\text{sc} + \xi)^2}{\xi (1+\kappa \xi)}}
\end{equation}
We see that in the limit $M=1$, $\kappa \rightarrow 0$ and $\chi \rightarrow 1$, and we recover the single bunch optimal efficiency and charge $q_\text{opt} = \mathcal{E}_0 / 2 h$ and $\eta_\text{opt} = \eta_\text{sc}$. 

For a given number of bunches $M$, the value of $q_\text{opt}$ determines a corresponding total macrobunch population $N_\text{opt}(M) = M q_\text{opt}(M)/e$. Let us assume that we want the average geometrical luminosity rate to match some reference value $\bar{\mathcal{L}}_\text{ref}$. Then for a round beam we require
\begin{equation}
    \bar{\mathcal{L}}_0(M) = \frac{N_\text{opt}(M)^2 f_\text{rep}}{2 \pi \sigma_0^2} = \bar{\mathcal{L}}_\text{ref} \ \ ;\ \sigma_0^2 = \epsilon_n \beta^* / \gamma\beta
    \label{eqn:constrainedlumi}
\end{equation}
where $\sigma_0$ is the incident radial beam size, $\beta^*$ is the beta function at the IP, $f_\text{rep}$ is the repetition rate of the laser, and $\epsilon_n$ is the radial normalized transverse emittance. Here $N_\text{opt}^2(M)$ highlights the fact that $N_\text{opt}$ is a function of bunch number $M$. 

\section{Parameter Optimization}
\label{sec:params}
In evaluating Eq. (\ref{eqn:constrainedlumi}), we may treat $\mathcal{E}_0$, $f_\text{rep}$, $\epsilon_n$, and $\lambda$ as free parameters to be varied within a reasonable range. Our choice of parameters must also be constrained by at least three additional concerns: (1) the peak field in the structure should not exceed the damage limit of the material, (2) the required wallplug power should be within reasonable bounds, and (3) the single-bunch charge $N/M$ should not exceed $10^5$ electrons/positrons to avoid space-charge blowup of the beam, as discussed in \cite{ALEGRO:2019}. We emphasize that we do not here attempt to determine a global optimum solution over the space of all free parameters. Rather, we evaluate Eq. (\ref{eqn:constrainedlumi}) with the above constraints in mind and present a self-consistent parameter set satisfying them which we use in subsequent sections to investigate microbunch luminosity enhancement. For the accelerator structure itself, we use the design of \cite{lin:2001} which is a silicon dioxide photonic crystal, having a cylindrical vacuum bore, leading to a cylindrical cavity mode, which is favorable to the case of a linear collider \cite{ALEGRO:2019}. The relevant structure parameters are given in Table \ref{tab:Lin}. 

\begin{table}[b]
\caption{\label{tab:Lin}%
Structure parameters corresponding to a cylindrical Bragg geometry \cite{mizrahi:2004}.
}
\begin{ruledtabular}
\begin{tabular}{lcc}
\textrm{Parameter}&
\textrm{Symbol}&
\textrm{Value}\\
\colrule
Normalized Group Velocity & $\beta_g$ & 0.5 \\
Laser Wavelength ($\mu$m) & $\lambda$ & 2  \\
Single-Stage Length (mm) & $L$ & 0.2 \\
Channel Radius ($\mu$m) & $a$ & 1.0  \\
Characteristic Impedance ($\Omega$) & $Z_C$ & 115  \\
Cherenkov Impedance ($\Omega$) & $Z_H$ & 240  \\
Fundamental Loss Factor ($\times 10^{20}$ V/m) & $k$ &  21.5 \\
Cherenkov Loss Factor ($\times 10^{20}$ V/m) & $h$ & $180$ \\
Damage Factor & $f_D$ & 2 \\
\end{tabular}
\end{ruledtabular}
\end{table}
To evaluate the damage limit, we use the empirical formula for laser induced damage fluence $F(\tau)$ of silicon dioxide, as reported originally in \cite{stuart:1995} and interpreted by \cite{hanuka:single:2018}. For the range of pulse lengths of interest (0.4 ps $ < \tau < 10$ ps), $F(\tau) \approx (2.51\ \text{J/cm}^2/\text{s}^{1/4}) \tau^{1/4}$. The corresponding peak field for a Gaussian mode profile is 
\begin{equation}
\label{eqn:Eth}
E_\text{th} = \sqrt{  \frac{2 Z_0 \epsilon_r F(\tau) } {\tau}}
\end{equation}
where $Z_0$ is the impedance of free space and $\epsilon_r = 2.13$ is the dielectric constant of silicon dioxide. We take $E_\text{th}$ to correspond to the peak field in the structure $E_\text{peak}$ which is related to the maximum axial field amplitude $\mathcal{E}_\text{max}$ by the so-called "damage factor" $f_D = E_\text{peak}/\mathcal{E}_\text{max}$. We equate the maximum field tolerated by the structure to the value $E_\text{th}$ above evaluated at the laser pulse duration $\tau_P$ and thereby obtain the threshold value $\mathcal{E}_\text{max} = E_\text{th}(\tau_P) / f_D$, which the axial field must not exceed in order to avoid damage to the structure. For the cylindrical Bragg geometry of Table \ref{tab:Lin}, $f_D \approx 2$. 

The wall-plug power is given by
\begin{equation}
    P_\text{wall} = \frac{P_\text{beam}}{\eta_L\ \eta_c\ \eta_\text{opt}} \ ; \ P_\text{beam} = N_\text{opt} \gamma m c^2 f_\text{rep}
\end{equation}
where $\eta_L$ is the laser wall-plug efficiency, $\eta_c$ is the optical coupling efficiency to the structure, $\eta_\text{opt}$ is the optimized coupling efficiency of the field to the beam as given by Eq. (\ref{eqn:efficiency}), and $P_\text{beam}$ is the beam power. Thulium fiber lasers have demonstrated $> 50\%$ wall plug efficiency and efficiencies as high as 82\% are theoretically possible \cite{moulton:2009}. For our calculations, we take $\eta_L = 75\%$. Since the optical coupling of the laser to the structure must be repeated many times (once for each stage of acceleration), optical losses must be carefully controlled, with coupling efficiency close to 100\%. We here assume 10\% optical coupling losses or $\eta_c = 90\%$.

In Table \ref{tab:optM} we list a self-consistent set of accelerator parameters satisfying the constraints imposed above on damage limits, power consumption, and microbunch charge. We note that these are similar to the parameters from \cite{england:rmp2014} in terms of assumed bunch charge, gradient, and efficiency. The top portion of the table represents starting parameters, and the bottom portion are the calculated results arising from solution of Eq. (\ref{eqn:constrainedlumi}).

\begin{table}[t]
\caption{\label{tab:optM}%
Parameters and bunch number $M_\text{opt}$ for matching to the reference luminosity $\bar{\mathcal{L}}_\text{ref}$ in Eq. (\ref{eqn:constrainedlumi}). 
}
\begin{ruledtabular}
\begin{tabular}{lcc}
\textrm{Parameter}&
\textrm{Symbol}&
\textrm{Value}\\
\colrule
Beam Energy (GeV) & $\gamma m c^2$ & 500  \\
Field Intensity (GV/m) & $\mathcal{E}_0$ & 1.1  \\
Train Repetition Rate (MHz) & $f_\text{rep}$ & 168  \\
IP Normalized Emittance (nm) & $\epsilon_n$ & 4.65 \\
IP Spot Size (nm) & $\sigma_0$ & 0.76 \\
Laser Pulse Duration (ps) & $\tau_P$ & 1.3 \\
Laser Wallplug Efficiency & $\eta_L$ & 0.75 \\
Optical Copuling Efficiency & $\eta_c$ & 0.90 \\
\colrule
Wake Correction (GV/m) & $\mathcal{E}_0(1+\delta)$ & 2.51  \\
Field Threshold (GV/m) & $\mathcal{E}_\text{max}$ & 2.85  \\
Unloaded Gradient (GeV/m) & $\langle G_0 \rangle$ & 1.80 \\
Loaded Gradient (GeV/m) & $G_L$ & 1.04 \\
Optimal Number of Bunches & $M_\text{opt}$ & 100  \\
Microbunch Population ($\times 10^4$) & $q_\text{opt}/e$ & 2.05 \\
Bunch Train Population ($\times 10^6$) & $N_\text{opt}$ & 2.05  \\
IP Beta Function ($\mu$m) & $\beta^*$ & 121 \\
Normalized Plasma Frequency & $\Omega^2$ & 0.042  \\
Normalized Emittance & $\Theta^2$ & 0.00027 \\
Coupling Efficiency & $\eta_\text{opt}$ & 39\%  \\
Wallplug Power (MW) & $P_\text{wall}$ & 103  \\
\end{tabular}
\end{ruledtabular}
\end{table}

\begin{figure}
\begin{center}
\includegraphics[height=0.58\textheight]{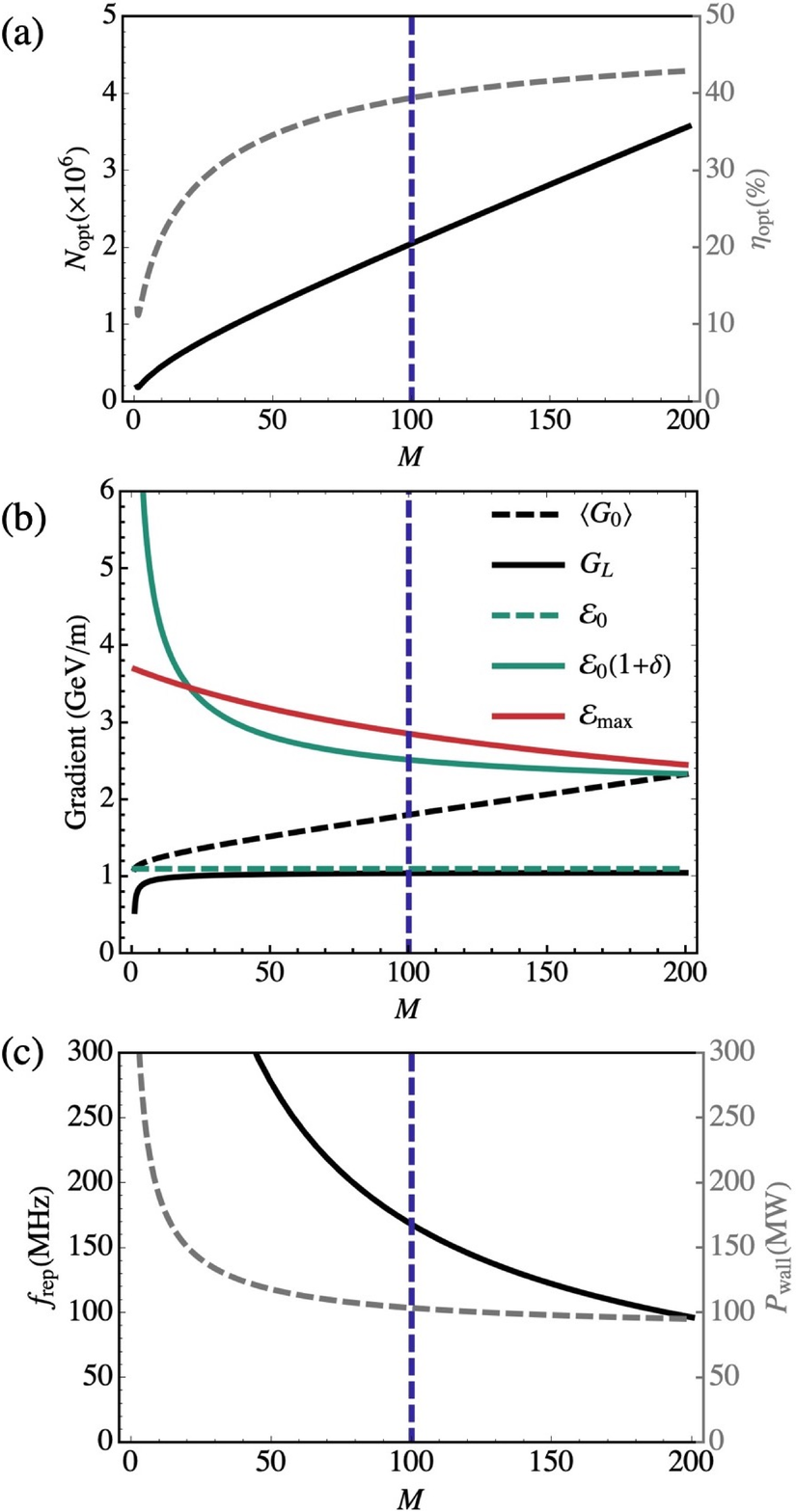}
\end{center}
\caption{Constrained parameters from Eq. (\ref{eqn:constrainedlumi}) as function of number $M$ of bunches, showing (a) macrobunch population (black) and optimal efficiency (gray); (b) corresponding field intensity parameters; and (c) corresponding laser repetition rate (black) and wallplug power (gray). All points on the curves are matched to $\bar{\mathcal{L}}_\text{ref}$ = $2\times10^{34} \text{cm}^{-2} \text{s}^{-1}$. Vertical blue dashed lines mark the value $M = 100$ corresponding to the parameters of Table \ref{tab:optM}.}\label{fig:params}
 \end{figure}

In Fig. \ref{fig:params}(a) we plot the constrained luminosity given by Eq. (\ref{eqn:constrainedlumi}) as a function of $M$ for the test case in Table \ref{tab:optM}. The horizontal orange line corresponds to the ILC reference $\bar{\mathcal{L}}_\text{ref}$ = $2 \times 10^{34}$ cm$^{-2}$ s$^{-1}$, and the vertical dashed line marks the intersection point with the luminosity curves. These points indicate the corresponding number of bunches $M_\text{opt}$ that satisfy \textit{both} equalities in Eq. (\ref{eqn:constrainedlumi}). The same points marked in Fig. \ref{fig:params}(b) and (c) indicate the corresponding values for optimal number of particles $N_\text{opt}$ and the resulting averaged unloaded and loaded gradients $\langle G_0 \rangle$ and $G_L$. Although higher luminosities are indicated in Fig. \ref{fig:params}(a) for $M>M_\text{opt}$ for either case, these higher luminosity values do not necessarily satisfy the imposed constraints on damage limits, wallplug power, and microbunch charge. The parameters in Table \ref{tab:optM} are summarized by the two dimensionless quantities $\Omega$ and $\Theta$ which are the normalized plasma frequency and emittance parameters to be explained in Section \ref{sec:eqsofmotion}. These two quantities will represent the primary input variables for the beam dynamics equations. 

\section{Luminosity Enhancement}
\label{sec:lumi}
The single-collision geometrical luminosity $\mathcal{L}$ for a pair of beams with number density distributions $\rho_1 (\textbf{r},t)$, $\rho_2 (\textbf{r},t)$ and velocities $\pmb{\beta}_1$, $\pmb{\beta}_2$ (normalized by speed of light $c$) may be written in relativistically invariant form \cite{Furman:Moller:2003}

\begin{equation}
\label{eqn:lumigeneral}
    \mathcal{L} = \int S(\textbf{r},t) d^3r dt
\end{equation}
where $S = \mathcal{K} \rho_1 \rho_2$ and $\mathcal{K}$ is the \textit{Moller luminosity factor}. 
\begin{equation}
\label{eqn:Moller}
    \mathcal{K} = c \sqrt{|\pmb{\beta}_1 - \pmb{\beta}_2|^2 - |\pmb{\beta}_1 \times \pmb{\beta}_2|^2}
\end{equation}
For counter-propagating ultra-relativistic Gaussian beams, $\mathcal{K} = 2 c$ and Eq. (\ref{eqn:lumigeneral}) yields the familiar form
\begin{equation}
\label{eqn:lumi}
    \mathcal{L} = \frac{N_1 N_2}{4 \pi \frac{1}{2}[(\sigma_{1x}^2+\sigma_{2x}^2)(\sigma_{1y}^2+\sigma_{2y}^2)]^{1/2}}
\end{equation}
where $N_1$ and $N_2$ are number of particles in beams 1 and 2 and $\sigma_{1x}$, $\sigma_{2x}$, $\sigma_{1y}$, $\sigma_{2y}$ are the Gaussian widths at the IP. The corresponding average luminosity per unit time is then $\bar{\mathcal{L}} = \mathcal{L} f_\text{rep}$, where $f_\text{rep}$ is the bunch repetition rate. If the two colliding beams each consist of a train of $M$ microbunches with $n = N/M$ particles per microbunch, it has previously been assumed for DLA like scenarios \cite{england:rmp2014}, where the full train of microbunches is generated by a single sub-picosecond laser pulse, that $N \rightarrow M n$ in Eq. (\ref{eqn:lumiround}). Since the microbunches are closely spaced at the laser wavelength $\lambda$ within each macrobunch, the constituent bunch trains effectively collide in unison. This differs from bunch trains in conventional RF accelerator scenarios, where instead one makes the replacements $N \rightarrow n$ and $f_\text{rep} \rightarrow M f_\text{rep}$ and so the average luminosity scales as $M n^2$ rather than $(M n)^2$. 

\begin{figure}
\begin{center}
\includegraphics[height=0.32\textheight]{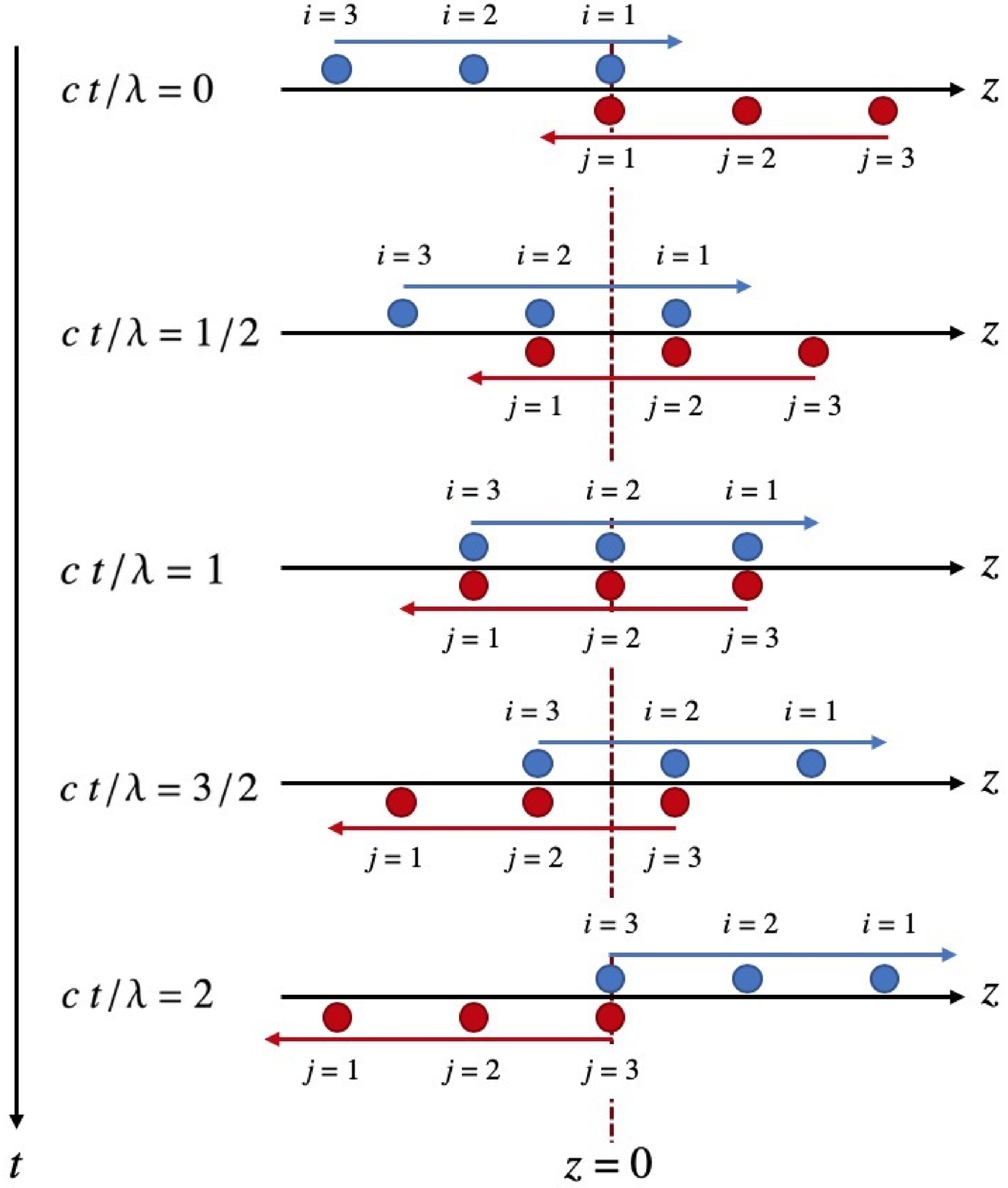}
\end{center}
\caption{Schematic of the $e^+ e^-$ collision process with $M = 3$ microbunches per train. At $t = 0$ the first microbunches of each train collide. For an arbitrary $M$, the full duration of the interactions is $\Delta t = (M-1)\lambda/c$ with a total of $M^2$ microbunch crossings. }\label{fig:collision}
 \end{figure}

The luminosity and corresponding enhancement for the collision of microbunch $i$ of one beam with microbunch $j$ of the other beam may be written, following from Eq. (\ref{eqn:lumi}),
\begin{equation}
\label{eqn:lumiround}
    \mathcal{L}_{ij} \equiv \frac{n^2}{4 \pi} \frac{4}{ (\sigma_i^2 + \sigma_j^2)} \ ; \ \mathcal{H}_{ij} \equiv \frac{\mathcal{L}_{ij}}{\mathcal{L}_0}= \frac{2 \sigma_0^2}{\sigma_i^2 + \sigma_j^2}    \end{equation}
where we assume initially round beams with the same number of particles ($N_1 = N_2 = N = M n$) but different radial Gaussian beam sizes ($\sigma_{ix} = \sigma_{iy} = \sigma_i/\sqrt{2}$, $\sigma_{jx} = \sigma_{jy} = \sigma_j/\sqrt{2}$). During each crossing, space charge attraction will tend to reduce the transverse sizes of the two beams. If both beams are assumed initially identical and round (but Gaussian and with opposite charge) in the absence of the other, with nominal IP spot size $\sigma_0$, then the corresponding geometrical luminosity is $\mathcal{L}_0 = N^2/2\pi \sigma_0^2$. The enhanced luminosity, accounting for the possibility of different  sizes of the crossing beams, is of the form of Eq. (\ref{eqn:lumiround}), and so the multi-bunch luminosity and enhancement factor may be written
\begin{equation}
\label{eqn:lumifull}
    \mathcal{L} = \sum_{i,j = 1}^M \mathcal{L}_{ij} \ \  ; \ 
    \mathcal{H}_D = \frac{\mathcal{L}}{\mathcal{L}_0} = \frac{1}{M^2} \sum_{i,j = 1}^M \frac{2 \sigma_0^2}{ (\sigma_i^2 + \sigma_j^2)}
\end{equation}
where $\sigma_0$ is the nominal radial RMS IP spot size of either beam in the absence of the other and $\sigma_i$ and $\sigma_j$ are assumed to be evaluated within the sum at the corresponding time at which bunches $i$ and $j$ cross. Figure \ref{fig:collision} illustrates the schematics of the collision process for the simple example of three microbunches ($M = 3$) in each train. We note by virtue of Eq. (\ref{eqn:lumigeneral}) that in principle the crossing of each microbunch should be integrated in time, since the spot sizes are functions of $t$. However, due to the  small bunch charges and short (sub-optical) time scales of each microbunch, the resulting correction is small. In the following section, we derive an analytical formulation for the transverse dynamics for colliding microbunch trains to numerically evaluate Eqs. (\ref{eqn:lumifull}).

\section{Equations of Motion - Constant Energy}
\label{sec:eqsofmotion}

The theory of beam-beam interaction at the IP of a collider has been extensively developed by prior authors \cite{hollebeek:1980,chen:1988,yokoya:1990,chen:USPAS:1987}. We extend these results to consider trains of colliding electron and positron bunches. We initially consider the case of constant energy, to be followed in the subsequent section by inclusion of the beamstrahlung energy losses due to the radiation reaction force (RRF). We begin by considering a single particle interacting with the transverse fields produced by the counter-propagating train of $M$ microbunches in the other beam, which we may represent by the number density function 
\begin{equation}
\label{eqn:rho}
    \rho(\textbf{r},t) = \frac{1}{M} \sum_{j=1}^M F_j(\textbf{r}_\perp) G_j(z,t)
\end{equation}
where $\textbf{r}_\perp = r \hat{\textbf{r}} = x \hat{\textbf{x}} + y \hat{\textbf{y}}$ is the radial coordinate and $F_j$, $G_j$ denote the transverse and longitudinal distribution functions for a round beam
\begin{align}
\label{eqn:F}
    &F_j(\textbf{r}_\perp) \equiv 2 [\mathcal{G}(r,\sigma_j)]^2,\\
\label{eqn:G}
    &G_j(z,t) \equiv \mathcal{G}(z+\beta ct - (j-1) \lambda, \sigma_z) .
\end{align}
Here $\mathcal{G}(\xi,\sigma) \equiv [1/\sqrt{2 \pi} \sigma] \exp [-\xi^2/(2 \sigma^2)]$ is a generic Gaussian, and $\sigma_j$, $\sigma_i$ are assumed to be time-dependent functions, while the length of the pulses $\sigma_z$ is assumed constant in time and the same for all bunches. A particle of charge $e$ sees a time-dependent transverse focusing force due to $\rho(\textbf{r},t)$ of the counter-propagating beam, which may be written (see Derivation in Appendix A)
\begin{equation}
    \textbf{F}_\perp(\textbf{r}_\perp,z,t) = -\gamma_0 m K^2(z,t)\ \textbf{r}_\perp
\label{eqn:Fperp} \ ,
\end{equation}
where $m$ is the electron/positron mass, $\gamma_0$ is the initial particle Lorentz energy factor, $K^2$ is the linear focusing term,
\begin{equation}
\label{eqn:K2}
    K^2(z,t) \equiv \frac{c}{\lambda} \frac{\Omega^2}{M} \sum_{j=1}^M \frac{\sigma_0^2}{\sigma_j^2}\ \mathcal{G} \left(\frac{t}{2}+\frac{z-(j-1)\lambda}{2 c},\sigma_t \right)
\end{equation}
and we define the normalized plasma frequency
\begin{equation}
\label{eqn:omega}
\Omega^2 \equiv \frac{2 N r_e \lambda}{\gamma_0 \sigma_0^2} \ .
\end{equation}
Here $r_e$ is the classical electron radius and $\sigma_t \equiv \sigma_z/2c$ is the interaction time of the bunches. The factor of $2$ here is due to the fact that in the lab frame, the beams have a relative velocity of $2c$. We have also made use of the fact that the generic Gaussian function satisfies the transformation property $\mathcal{G}(\xi,\sigma) = (1/a) \mathcal{G}(a \xi, a \sigma)$. Invoking the relativistic form of Newton's second law, $\partial_t (\gamma m \dot{\textbf{r}}_\perp) = \textbf{F}_\perp$, we thus obtain for the case of constant energy ($\gamma = \gamma_0$) the equations of motion
\begin{equation}
\label{eqn:rdot}
    \ddot{\textbf{r}}_\perp+K_i^2 \textbf{r}_\perp = 0 \ ,
\end{equation}
where overhead dots denote $t$ derivatives and 
\begin{equation}
\label{eqn:K2}
    K_i^2 \equiv \frac{c}{\lambda} \frac{\Omega^2}{M} \sum_{j=1}^M \frac{\sigma_0^2}{\sigma_j^2}\ \mathcal{G}\left(t-\frac{\lambda}{c} \left[\frac{i+j}{2}-1 \right], \sigma_t \right) \ .
\end{equation}
We note that $K_i$ is merely $K(z,t)$ evaluated at $z = c t - (i-1) \lambda$, thereby taking the test particle to lie in the $i$'th microbunch of the incident beam. To simplify future calculations, let us define the normalized time coordinate $u \equiv c t / \lambda$ and cast Eqs. (\ref{eqn:rdot},\ref{eqn:K2}) into the forms
\begin{equation}
\label{eqn:motion_u}
\textbf{r}_\perp'' + \hat{K}_i^2 \textbf{r}_\perp = 0
\end{equation}
\begin{equation}
\label{eqn:defs_K}
\hat{K}_i^2 \equiv \frac{\Omega^2}{M} \sum_{j=1}^M \frac{\sigma_0^2}{\sigma_j^2}\ \mathcal{G}(u-u_{ij},\sigma_u)\\
\end{equation}
where $u_{ij} \equiv (i+j)/2-1$ is the crossing time of bunches $i$ and $j$, and $\sigma_u \equiv \sigma_t c/\lambda$ is the microbunch duration in our normalized units. By analogy with the usual equation \cite{rosenz:1997} for the evolution of the transverse beam size under the action of a linear focusing force, we can write the corresponding envelope equation
\begin{equation}
\label{eqn:envelope}
\bar{\sigma}_i'' + \hat{K}_i^2 \bar{\sigma}_i - \frac{\Theta^2}{\bar{\sigma}_i^3} = 0 \ \ ; \ \ i = 1, 2, ..., M
\end{equation}
with the associations
\begin{equation}
    \label{eqn:defs_theta}
\Theta \equiv \frac{\lambda \epsilon_0}{\sigma_0^2} \ \ ; \ \ \bar{\sigma}_i \equiv \frac{\sigma_i}{\sigma_0}
\end{equation}
where $\epsilon_0$ is the initial transverse RMS emittance. Eqs. (\ref{eqn:envelope}) represent $M$ equations in the $M$ unknown functions $\bar{\sigma}_i(u)$. Due to the symmetry of the problem, the solutions for $\bar{\sigma}_i$ apply to both the electron and positron beam. This is referred to as a strong-strong interaction since the dynamics of both species vary in the course of the interaction. We can associate the plasma frequency appearing in this equation with $\Omega^2 = D/(M-1)$ where $D = N r_e \ell / (\gamma \sigma_0^2)$ is the disruption parameter of the macrobunch of length $\ell = (M-1) \lambda$. The dimensionless quantity $\Theta$ is a normalized angular momentum term which gives rise to the transverse beam emittance. Its definition in Eq. (\ref{eqn:defs_theta}) ensures that if the independent variable in Eq. (\ref{eqn:envelope}) is cast into units of path length $s = u \lambda = c t$ that the $\Theta^2$ term assumes the usual form of the emittance pressure $\epsilon_0^2 / \sigma^3$. Obviously this formulation is only valid close to the IP, and it suffices to set the initial conditions $\bar{\sigma}_i(0) = 1$ and $\bar{\sigma}_i'(0)=0$. Evidently, in the framework of this formulation, subject to the assumption that initially all the microbunches are identical, there are three degrees of freedom: the normalized emittance $\Theta$, plasma frequency $\Omega$ and number of microbunches $M$.

\section{Equations of Motion with RRF}
\label{sec:RRFeqns}

We now extend the theory to include the radiation reaction force (RRF) and thus the particle energy loss. Our analysis is limited to the classical regime under the assumption that the beamstrahlung parameter $\Upsilon \ll 1$ for cases of interest. The correctness of this assumption will be shown later. The relativistic theory of radiation reaction of classical point electrons derived by Dirac has the following Lorentz invariant 4-vector form:

\begin{equation}
    \frac{dp_\mu}{d\tau} = \mathcal{F}_\mu^\text{rad} + \mathcal{F}_\mu^\text{ext}
    \label{eqn:LAD}
\end{equation}
where $\tau$ is the proper time, $p_\mu$ is the 4-vector momentum, and $\mathcal{F}_\mu^\text{rad}$ and $\mathcal{F}_\mu^\text{ext}$ are the Lorentz invariant forms of the radiation reaction force and the externally applied force respectively. By using the requirement that any force must satisfy $\mathcal{F}_\mu p^\mu = 0$, the following form for the radiation force in the $(1,-1,-1,-1)$ metric may be written:
\begin{equation}
\mathcal{F}_\mu^\text{rad} = \frac{2}{3} \frac{r_e}{c} \left[\frac{d^2p_\mu}{d\tau^2}+\frac{p_\mu}{(mc)^2} \left(\frac{dp_\nu}{d\tau} \frac{dp^\nu}{d\tau}\right) \right]
\label{eqn:Frad}
\end{equation}
where $r_e$ is the classical electron radius. Defining the normalized momenta $q^\nu \equiv p^\nu / (m c) = (\gamma, \textbf{q})$, where $\textbf{q} \equiv \gamma \pmb{\beta}$, Eq. (\ref{eqn:LAD}) may be expressed to good approximation by (see derivation in Appendix \ref{appendixD})

\begin{align}
    \gamma' &\simeq \pmb{\beta} \cdot \bar{\textbf{F}}_\perp - \alpha_e |\textbf{q}'|^2 |\bar{\textbf{F}}_\perp|^2\\
    \textbf{q}' &\simeq \bar{\textbf{F}}_\perp - \alpha_e \gamma \textbf{q} |\bar{\textbf{F}}_\perp|^2
    \label{eqn:RRFeqnmotion}
\end{align}
where primes denote derivatives with respect to the  normalized time $u = c t/\lambda$, $\bar{\textbf{F}} \equiv \lambda \textbf{F} / (m c^2)$ is the normalized external force and $\alpha_e \equiv c \tau_e / \lambda$, with $\tau_e \equiv 2 r_e / (3 c)$. Noting by the arguments of the prior section that our form for the external force is $\bar{\textbf{F}}_\perp=-(\gamma_0/\lambda) \hat{K}_i^2 \textbf{r}_\perp$, these equations of motion lead to the following modified envelope equations for the $i$'th microbunch:

\begin{equation}
\begin{aligned}
   \bar{\sigma_i}'' + \frac{1}{\bar{\gamma}_i}\hat{K}_i^2 \bar{\sigma}_i-\frac{\Theta^2}{\bar{\sigma}_i^3}+\frac{\bar{\gamma}_i'}{\bar{\gamma}_i} \bar{\sigma}_i' + &\bar{\alpha}_e \hat{K}_i^4 \bar{\sigma}_i' \bar{\sigma}_i^2 \bar{\gamma}_i = 0 \\
   \bar{\gamma}_i' + \bar{\alpha}_e \hat{K}_i^4 \bar{\sigma}_i^2 \bar{\gamma}_i^2 &= 0
   \label{eqn:RRFenvelope}
\end{aligned}
\end{equation}
where we define the additional dimensionless parameters
\begin{equation}
    \bar{\alpha}_e \equiv \alpha_e \gamma_0^2 \left(\frac{\sigma_0}{\lambda}\right)^2  \ \ ; \ \ \bar{\gamma}_i \equiv \gamma_i / \gamma_0
\end{equation}
Equations (\ref{eqn:RRFenvelope}), derived in detail in Appendix \ref{appendixE}, represent $2 M$ equations in $2M$ unknowns ($\bar{\sigma}_i$ and $\bar{\gamma}_i$) subject to the initial conditions $\bar{\sigma}_i(0) = 1$, $\bar{\sigma_i}'(0)=0$, and $\bar{\gamma}_i(0)=1$. The primary assumptions underlying these equations are: (1) that the interaction is classical ($\Upsilon \ll 1$), (2) that the external beam-beam force is linear in $\textbf{r}_\perp$, and (3) the approximation of Landau and Lifshitz, whereby momentum derivatives appearing in $\mathcal{F}_\mu^\text{rad}$ may to first order be approximated by their usual relativistic forms in the absence of RRF. The quantities multiplied by $\bar{\alpha}_e$ represent the RRF contributions. We see further that if these terms are neglected in the limit $\bar{\gamma}_i' \rightarrow 0$, Eq. (\ref{eqn:RRFenvelope}) reduce to the form of Eq. (\ref{eqn:envelope}) for constant energy $\gamma_i = \gamma_0$. 

An additional observation is that the normalized emittance under the action of the RRF is not a constant of motion, as it is in the case of simple acceleration under linear focusing, but rather varies in proportion to $\gamma$. By virtue of this observation, a third set of equations is implicit in Eq. (\ref{eqn:RRFenvelope}) of the form $\epsilon_{n,i}(u) = \gamma_i(u) \epsilon_0$, where $\epsilon_{n,i}$ is the normalized emittance of the $i$'th microbunch and $\epsilon_0$ is the initial RMS (un-normalized) emittance. This result is also derived and explained in Appendix \ref{appendixE}. To our knowledge this effect has not been previously noted. 

\section{Energy Loss}
\label{sec:upsilon}

The beamstrahlung parameter is a measure of the degree to which an electromagnetic interaction may be described classically, given by the Lorentz invariant
\begin{equation}
    \Upsilon = \frac{e \hbar}{m^3 c^4}[p_\mu F^{\mu \xi} p^\nu F_{\xi \nu}]^{1/2}
\end{equation}
where $p_\mu$ is the 4-momentum and $F_{\mu \nu}$ is the mean energy-momentum stress tensor of the beam field. Letting $E_c = (m^2 c^3)/(e \hbar)$ denote the Schwinger critical field, we find for a purely transverse force $\textbf{F}_\perp$ that
\begin{equation}
\Upsilon = \gamma \frac{|\textbf{F}_\perp|}{e E_c}
\end{equation}
Noting that the transverse force in the lab frame is given by $\textbf{F}_\perp = -\gamma_0 m K_i^2 \textbf{r}_\perp$, averaging over the transverse beam distribution $F_i(\textbf{r}_\perp)$, and defining the normalized transverse focusing for interaction $(i,j)$ by 
\begin{equation}
    \hat{K}_{ij}^2 \equiv \frac{\Omega^2}{M} \frac{1}{\bar{\sigma}_j^2} \mathcal{G}(u -u_{ij},\sigma_u)
\end{equation}
where $u_{ij} \equiv (i+j)/2-1$, we obtain the following form for the spatially averaged beamstrahlung parameter for interaction $(i,j)$:
\begin{equation}
   \Upsilon_{ij}(u) = \bar{\gamma}_i \bar{\sigma}_i \frac{\lambdabar_c}{\lambda} \frac{\sqrt{\pi}}{2}\left( \frac{\sigma_0}{\lambda} \right) \gamma_0^2 \hat{K}_{ij}^2
\end{equation}
where $\lambdabar_c = \hbar/(m c)$ is the reduced Compton wavelength. Let us define the time averaged beamstrahlung parameter by integrating over $u$:
\begin{equation}
    \langle \Upsilon_{ij} \rangle \equiv \frac{5}{6} \frac{1}{\sqrt{2 \pi} 2 \sigma_u} \int_{-\infty}^{\infty} \Upsilon_{ij}(u) du
\end{equation}
Here we include the extra factor of $5/6$ which appears in some formulations, noting that this factor is merely for convenience and has no physical significance. For short bunches we may approximate the function $\mathcal{G}(u,\sigma_u)$ by a Dirac delta function and simply evaluate the integrand at the time $u_{ij}$ when bunches $i$ and $j$ are overlapped. We thus obtain
\begin{equation}
    \langle \Upsilon_{ij} \rangle \simeq \frac{5}{6} \frac{\gamma_0^2}{4 \sqrt{2} \sigma_u} \frac{\lambdabar_c}{\lambda} \left( \frac{\sigma_0}{\lambda} \right) \frac{\Omega^2}{M} \frac{\bar{\gamma}_i \bar{\sigma}_i}{\bar{\sigma}_j^2} \bigg|_{u=u_{ij}}
    \label{eqn:Upsilonij}
\end{equation}
For comparison with the usual form of this parameter, we may recast the above relation into the equivalent form
\begin{equation}
      \langle \Upsilon_{ij} \rangle \simeq \Upsilon_\text{sc} \times \frac{\bar{\gamma_i} \bar{\sigma}_i}{\bar{\sigma}_j^2} \bigg|_{u=u_{ij}} \ ; \ \ \Upsilon_\text{sc} \equiv  \frac{5}{6}\frac{\gamma_0 r_e^2 (N/M)}{\alpha \sigma_z \sqrt{2} \sigma_0}
\label{eqn:Upsilonsc}
\end{equation}
We recognize the expression for $\Upsilon_\text{sc}$ as the familiar form for the single-crossing beamstrahlung parameter for two bunches of equal size containing $N/M$ particles. Since $\sigma_0$ is here the \textit{radial} RMS beam size, for comparison with the usual formula we would make the replacement $\sigma_0 \rightarrow (\sigma_x+\sigma_y)/\sqrt{2}$. We further see that in the limit of a single crossing ($M \rightarrow 1$) that the correction term is unity at $u = 0$ and hence $\langle \Upsilon_{ij} \rangle \rightarrow \Upsilon_\text{sc}$ as expected. The average beamstrahlung parameter may be obtained by summing over $i$ and $j$:
\begin{equation}
    \langle \Upsilon \rangle = \frac{1}{M^2} \sum_{i,j = 1}^{M} \langle \Upsilon_{ij} \rangle \ .
    \label{eqn:Upsilonavg}
\end{equation}

The corresponding fractional energy loss for the $(i,j)$ interaction may be obtained by integrating $\bar{\gamma}'$ in Eq. (\ref{eqn:RRFenvelope}), yielding
\begin{equation}
 \delta\bar{\gamma}_{ij} =
    \frac{\bar{\alpha}_e}{2 \sqrt{\pi} \sigma_u} \frac{\Omega^4}{M^2} \frac{\bar{\gamma_i}^2 \bar{\sigma}_i^2}{\bar{\sigma}_j^4} \bigg|_{u=u_{ij}}
    \label{eqn:dgammaint} \ .
\end{equation}
Here we use the fact that for a Gaussian function $\mathcal{G}$, $\int f(\xi) [\mathcal{G}(\xi-\xi_0, \sigma)]^2 d\xi \approx f(\xi_0)/(2\sqrt{\pi} \sigma)$. Combining Eqs. (\ref{eqn:dgammaint}) and (\ref{eqn:Upsilonij}), we obtain:
\begin{equation}
   \delta\bar{\gamma}_{ij} = \frac{192}{25 \sqrt{\pi}} \frac{\alpha}{\Gamma_0} \langle \Upsilon_{ij} \rangle^2 
   \label{eqn:dgammaij}
\end{equation}
where $\Gamma_0 \equiv \lambdabar_c \gamma_0 /\sigma_z$ and $\alpha$ is the fine structure constant. Eq. (\ref{eqn:dgammaij}) is correct to the same level of approximation as our envelope equations (\ref{eqn:RRFenvelope}) and thus should provide a close match to its predictions of energy loss of the bunches. The fractional energy loss of the entire bunch train is then given by
\begin{equation}
    \langle \delta\bar{\gamma} \rangle = \langle \frac{\delta \gamma}{\gamma_0} \rangle \simeq \frac{1}{M} \sum_{i,j = 1}^{M} \delta\bar{\gamma}_{ij}
    \label{eqn:deltagammaavg}
\end{equation}

We additionally note that because our form of the transverse force was truncated to lowest order in the transverse coordinate, the energy loss given by Eq. (\ref{eqn:deltagammaavg}) is actually an overestimate. The correction term $\mathcal{Z}_{ij} = (25 \pi / 288) \bar{Z}_{ij}$ is derived in Appendix \ref{appendixC}. Multipyling it with Eq. (\ref{eqn:dgammaij}), we obtain the \textit{reduced} energy loss 
\begin{equation}
\label{eqn:dgammareduced}
    \langle \delta\bar{\gamma} \rangle_\text{red} = \frac{2 \sqrt{\pi}}{3} \frac{\alpha}{\Gamma_0} \frac{1}{M} \sum_{i,j = 1}^{M}  \langle \Upsilon_{ij} \rangle^2 \bar{Z}_{ij}
\end{equation}
where
\begin{equation}
\bar{Z}_{ij} \equiv \frac{\mu^4}{\ln (2/\sqrt{3})} \ln [\frac{1+\mu^2}{\mu \sqrt{2+\mu^2}}] \ ; \ \mu \equiv \frac{\bar{\sigma}_j}{\bar{\sigma}_i}\bigg|_{u=u_{ij}}
\end{equation}
The results of Eqs. (\ref{eqn:RRFenvelope}) and (\ref{eqn:deltagammaavg}) therefore represent conservative estimates of the energy loss. We see that in the limit of a single bunch crossing ($M =1$) where the bunches are of equal size ($\mu = 1$) then $\langle \Upsilon_{ij} \rangle \rightarrow \Upsilon_\text{sc}$ and $\bar{Z}_{ij} \rightarrow 1$, and thus Eq. (\ref{eqn:dgammareduced}) reduces to
\begin{equation}
     \langle \delta\bar{\gamma} \rangle_\text{red} \rightarrow \frac{2 \sqrt{\pi}}{3} \frac{\alpha}{\Gamma_0} \Upsilon_{\text{sc}}^2
\end{equation}
which is the usual formula for single bunch energy loss.

\section{Numerical Results}
\label{sec:results}

We evaluate numerically the solutions to Eq. (\ref{eqn:RRFenvelope}) for some typical parameters consistent with a DLA based collider presented in Table \ref{tab:optM}. To solve the normalized envelope equations (\ref{eqn:RRFenvelope}), we require the two dimensionless input parameters $\Omega^2$=0.042 and $\Theta^2$ = 0.00027. For the normalized plasma frequency in both cases with $M = 100$, the disruption parameter is of order $D = M \Omega^2 \approx 4.2$. In the numerical evaluations that follow we keep the disruption as specified, keeping the simulation and its accuracy at a reasonable level but without affecting the physical process. 

\begin{figure}[tbh]
\begin{center}
\includegraphics[height=0.67\textheight]{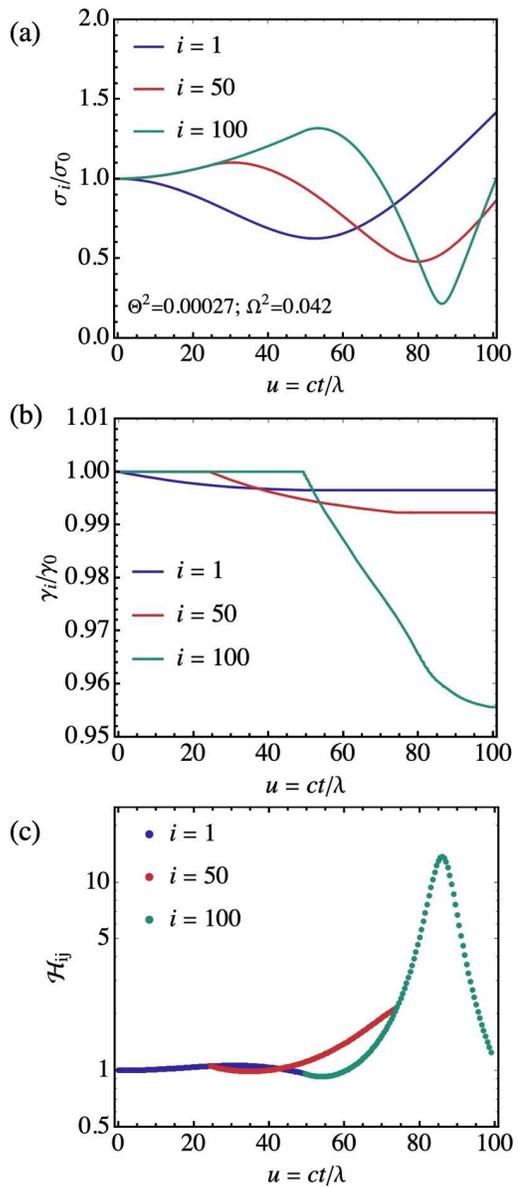}
\end{center}
\caption{Frames (a) and (b) show the normalized spot size $\bar{\sigma}_i = \sigma_i / \sigma_0$ and energy $\bar{\gamma}_i = \gamma_i / \gamma_0$ respectively from solving Eq. (\ref{eqn:RRFenvelope}) for three representative microbunches $i$ = 1, 50, 100 in a train of $M=100$ bunches. The calculations are performed in the time range $0 \leq t \leq (M-1)\lambda/c$. The parameters correspond to those of Table \ref{tab:optM} with $\Omega^2$ = 0.042 and $\Theta^2 = 2.7 \times 10^{-4}$. Part (c) shows the partial luminosity enhancement factor $\mathcal{H}_{ij}$ along each trajectory.}
\label{fig:enhancement}
 \end{figure}
 
 Let us start by examining the solutions for three microbunches in a bunch train with $M$ = 100 bunches. We consider the first microbunch $i = 1$, the central one $i = 50$ and the last one $i = 100$. In Figure \ref{fig:enhancement}(a) we present the normalized envelope of the three microbunches. According to this plot, the first microbunch pinches to half of its initial transverse size after a time $t = 50 \lambda/c$ which is half of the total interaction time $\Delta t = (M-1) \lambda/c$. Beyond this point the repelling angular momentum term $\Theta$ becomes dominant. As a result of the pinching effect, later bunches see a higher density in the counter-propagating beam and thus are focused more strongly, as shown by the envelopes of the middle ($i = 50$) and final ($i = 100$) bunches. 
 
The initial transverse emittance, here represented by the normalized variable $\Theta$, sets the minimum envelope of each microbunch. If we reduce it, the pinching is more pronounced. A lower $\Theta$ facilitates stronger pinching but a fraction of the microbunches expand back to beam sizes much larger than the initial condition of the envelope. In part (b) of Fig. \ref{fig:enhancement}, we plot the solutions for $\bar{\gamma}_i$ for each of the bunches, which show increasing energy loss with later bunches in the train, due to the increased deflection of the particle trajectories and hence increased RRF-induced energy loss. In Fig. \ref{fig:enhancement}(c) we plot the partial luminosity enhancement $\mathcal{H}_{ij}$ evaluated for each of the three values of $i$ at the crossing times $u = u_{ij} = (i+j)/2-1$. We see that another byproduct of the pinching effect is that the luminosity enhancement increases for later interactions.

\begin{figure}
\begin{center}
\includegraphics[height=0.67\textheight]{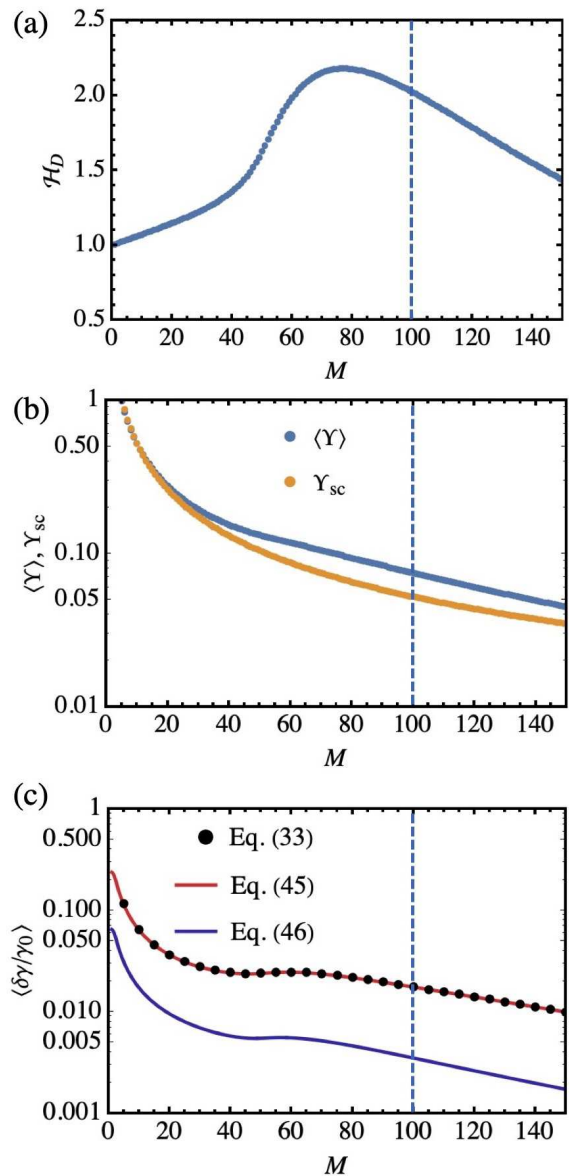}
\end{center}
\caption{Computed solutions for (a) luminosity enhancement, (b) beamstrahlung parameter, and (c) fractional energy loss of the entire bunch train for various number of bunches $M$ for the case $\Omega^2 = 0.042$, $\Theta^2 = 0.00027$. Note that the values of Table \ref{tab:optM} correspond to $M$ = 100 on these plots (vertical dashed blue markers). In part (c), numerical solutions of fractional energy loss (black dots) are shown for every 5th $M$ value to aid visualization. Including nonlinearity of the transverse force yields approximately a factor of 4 reduction in energy loss (blue curve).}
\label{fig:eloss}
\end{figure}

In Fig. \ref{fig:eloss}, we plot the luminosity enhancement, beamstrahlung parameter, and energy loss \textit{of the entire bunch train} as we vary the number of bunches from $M=1$ to $M=150$ while keeping fixed the other parameters of Table \ref{tab:optM}. Hence, the points on these curves correspond to the same geometrical (un-enhanced) luminosity $\mathcal{L}_0$, but with the macrobunch population $N$ distributed between different numbers of bunches. In Fig. \ref{fig:eloss}(a), the luminosity enhancement is seen to peak at around $M = 70$. We find that the peak of the enhancement scales approximately as $\mathcal{H}_D \propto \Theta^{-1} \propto (\epsilon/\sigma_0^2)^{-1}$. Beyond its peak value, $\mathcal{H}_D$ decreases monotonically with the number of microbunches. In fact, according to our calculations, the product $M \mathcal{H}_D^{(M \gg 1)}$ is independent of $M$, thus we conclude that roughly, $\mathcal{H}_D^{(D>5)} \propto D^{-1}$. In the range $1 \leq M \leq 30$, before the peak of $\mathcal{H}_D$, the enhancement factor has a weak linear dependence on the disruption parameter ($1 \leq D \leq 1.5$). This linear dependence is virtually independent of $\Theta$ and thus of the emittance. In Fig. \ref{fig:eloss}(b), the corresponding average beamstrahlung parameter $\langle \Upsilon \rangle$ from Eq. (\ref{eqn:Upsilonavg}) is plotted, and compared with the canonical single-crossing $\Upsilon_\text{sc}$ defined in Eq. (\ref{eqn:Upsilonij}).

\begin{table}[t]
\caption{\label{tab:summary}%
Summary of luminosity enhancement and corresponding energy loss for the two test cases of Table \ref{tab:optM}. 
}
\begin{ruledtabular}
\begin{tabular}{lcc}
\textrm{Parameter}&
\textrm{Symbol}&
\textrm{Value} \\
\colrule
Plasma Frequency & $\Omega^2$ & 0.042  \\
Angular Momentum & $\Theta^2$ & $0.00027$  \\
Optimal Number of Bunches & $M_\text{opt}$ & 100 \\
Luminosity Enhancement Factor & $\mathcal{H}_D$ & 2.0  \\
Geo. Luminosity ($10^{34}\text{cm}^{-2}\text{s}^{-1}$) & $\bar{\mathcal{L}}_0$ & 2  \\
Enh. Luminosity ($10^{34}\text{cm}^{-2}\text{s}^{-1}$) & $\mathcal{H}_D \bar{\mathcal{L}}_0$ & 4.0   \\
Avg Beamstrahlung Param & $\langle \Upsilon \rangle$ & 0.075 \\
Average Energy Loss & $\langle \delta\bar{\gamma} \rangle$ & 1.75 \%  \\
Reduced Energy Loss & $\langle \delta\bar{\gamma} \rangle_\text{red}$ & 0.35 \%  \\
\end{tabular}
\end{ruledtabular}
\end{table} 

In Table \ref{tab:summary} we report the corresponding energy loss values for the parameters of Table \ref{tab:optM}. The results support one of the fundamental assumptions of our analysis, which is that the beamstrahlung parameter $\langle \Upsilon \rangle \ll 1$, and therefore that a classical treatment is approprite. The total fractional energy loss $\langle \delta \bar{\gamma} \rangle$ is $1.75\%$, or $0.35\%$ with the inclusion of the reduction term $Z_{ij}$ that accounts for the nonlinear dependence of the external force on transverse coordinate, as in Eq. (\ref{eqn:dgammareduced}). This is a very important result since a factor of 2 in luminosity enhancement is by far more important than a fraction of one percent reduction of the particle energy. 

\section{Conclusions}
\label{sec:conclusions}

In this study, we have numerically calculated the luminosity enhancement and radiative energy loss for a 1 TeV center of mass $\text{e}^+\text{e}^-$ collider scenario with a set of self-consistent DLA beam parameters. We find that the optically microbunched DLA beam leads to a significant luminosity enhancement $\mathcal{H}_D$ due to the contraction (pinch) of the two counter-propagating trains of microbunches. We find that for a given plasma frequency and emittance, there is a particular value $M \approx 70$ where the luminosity enhancement $\mathcal{H}_D$ is maximal, and that this value is independent of the emittance, at least in the regime of interest. 

However, the optimal $\mathcal{H}_D$ value is inversely proportional to the emittance. For small number of microbunches in the train ($M < 30$) $\mathcal{H}_D$ varies linearly with $M$, while for larger values ($M > 70$), $\mathcal{H}_D$ scales inversely with $M$. For the considered parameter case, we determined an optimal number of bunches $M_\text{opt} = 100$ that matches the desired geometrical luminosity while satisfying constraints on beam loading efficiency, power consumption, and bunch charge. We find that there is a tradeoff between having $M_\text{opt}$ matched to the peak of the luminosity curve and the overall amplitude of this peak. Although the scenario of Table \ref{tab:optM} requires $M_\text{opt} = 100$ microbunches and so does not sit at the peak of the $\mathcal{H}_D$ curve, it also leads to lower required repetition rate, lower maximal field in the structure, lower single-bunch charge, higher efficiency, and reduced wallplug power consumption, making it generally favorable. 

We see that the overall beamstrahlung energy losses are 1.75\% for the unreduced case and 0.35\% for the reduced case (the latter taking into account the full transverse shape of the beam potential $\Psi$). We note that our analysis predicts similar values of the luminosity enhancement and beamstrahlung energy loss to prior estimates (e.g. Ref. \cite{england:rmp2014}), which are on the order of $\mathcal{H}_D = 2.7$ and $ \langle \delta\bar{\gamma} \rangle= 1\%$ when scaled to 1 TeV center of mass energy. This is in spite of the fact that here we treat the microbunches as distinct, whereas prior calculations assumed a ``washed out" microstructure, treating the entire bunch train as a single longer bunch. The small energy loss here is an important potential benefit of the DLA approach for a linear collider, particularly in the multi-TeV regime, where the energy losses for more conventional bunch formats can be in the tens of percents. 

\begin{acknowledgments}
This work was supported by the Israel Science Foundation (ISF), the Moore Foundation (GBMF4744), and the U.S. Department of Energy (DE-AC02-76SF00515). 
\end{acknowledgments}

\appendix

\section{Transverse Potential}
\label{appendixA}

\begin{figure}[h]
\begin{center}
\includegraphics[height=0.12\textheight]{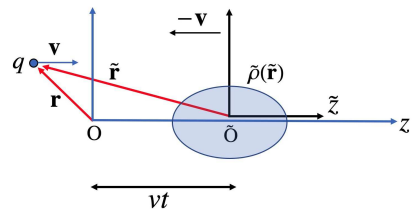}
\end{center}
\caption{Coordinate system for single particle of velocity $v$ colliding with the charge density function $\rho$ with velocity -$v$. }\label{fig:coordinates}
 \end{figure}

Let $\text{O}$ denote the lab frame and $\tilde{\text{O}}$ the frame of a counter-propagating charge distribution moving at relative velocity $v = -\beta c$ along the $z$-axis such that the origins of $\text{O}$ and $\tilde{\text{O}}$ overlap at time $t=0$. This is illustrated in Fig. \ref{fig:coordinates}.

Let \textbf{r} be the position of a test charge $q$ in the lab frame. Then its position in $\tilde{\text{O}}$ is obtained by transforming only the longitudinal component $\tilde{z} = \gamma (z+ v t)$. Hence, the gradient of the electrostatic potential in this frame is
\begin{equation}
    \tilde{\boldsymbol{\nabla}} \tilde{\Phi} = \boldsymbol{\nabla} \tilde{\Phi}(\textbf{r}_\perp, \tilde{z})\big|_{\tilde{z} = \gamma (z+\beta c t)}
\end{equation}
In \cite{yokoya:1990} the potential for a tri-Gaussian beam distribution is found to have the following form
\begin{equation}
    \tilde{\Phi}(\tilde{\textbf{r}}) = \frac{N e}{4 \pi \varepsilon_0} \mathcal{G}(\tilde{z},\tilde{\sigma}_z) \Psi(\tilde{\textbf{r}}_\perp)
\end{equation}
where $\mathcal{G}(\tilde{z},\tilde{\sigma}_z)$ is the Gaussian longitudinal charge distribution normalized to unit integral over $\tilde{z}$ and
\begin{equation}
   \Psi(\tilde{\textbf{r}}_\perp) = \int_0^\infty \frac{ \exp(-\frac{\tilde{x}^2}{2 \sigma_x^2+\xi}-\frac{\tilde{y}^2}{2 \sigma_y^2+\xi})} {\sqrt{(2 \sigma_x^2 + \xi)(2 \sigma_y^2+\xi)}} d\xi
\end{equation}
In \cite{herr:2003} the following approximate form for the gradient of this potential is derived for a round beam ($\sigma \equiv \sqrt{2} \sigma_x = \sqrt{2} \sigma_y$):
\begin{equation}
\label{eqn:potential}
    \boldsymbol{\nabla}_\perp \Psi = \frac{2 \textbf{r}_\perp}{r^2} \left[1-\exp(\frac{r^2}{\sigma^2})\right]\approx \frac{2 \textbf{r}_\perp}{\sigma^2}
\end{equation}
In the second step, we expand the term in square brackets to lowest order in $r/ \sigma^2$ and we remove the tildes, since the transverse coordinates are not affected by the Lorentz transformation and so are the same in both frames. 

\section{The External Force}
\label{appendixB}

Referencing the coordinates of Appendix \ref{appendixA}, consider now a static charge density $\tilde{\rho}$ that is stationary in frame $\tilde{\text{O}}$. In this frame, the charge density $\tilde{\rho}$ gives rise to the electrostatic potential $\tilde{\Psi}$ with corresponding electric and magnetic fields
\begin{equation}
    \tilde{\textbf{E}} = -\tilde{\boldsymbol{\nabla}}\tilde{\Phi} \ \ ; \  \ \tilde{\textbf{B}} = 0.
\end{equation}
Transforming from the $\tilde{\text{O}}$ to O frame, we have that $\textbf{E}_\perp = \gamma \tilde{\textbf{E}}_\perp$ and $\textbf{B}_\perp = \gamma \beta (\tilde{E}_y\hat{\textbf{x}}- \tilde{E}_x\hat{\textbf{y}})$. The resulting transverse Lorentz force on the particle of charge $-e$ in the lab frame with normalized velocity $\pmb{\beta} = \beta \hat{\textbf{z}}$ is thus
\begin{equation}
    \textbf{F}_\perp = -e (\textbf{E} + \pmb{\beta} \times \textbf{B})_\perp = -e \gamma (1+\beta^2) \tilde{\boldsymbol{\nabla}}_\perp \tilde{\Phi}
\end{equation}
Regarding the potential as a 2D function with separable longitudinal part, it was shown in Appendix \ref{appendixA} that
\begin{equation}
    \tilde{\boldsymbol{\nabla}}_\perp \tilde{\Phi} = \frac{N e}{4 \pi \varepsilon_0} \mathcal{G}(\tilde{z},\tilde{\sigma}_z) \boldsymbol{\nabla}_\perp \Psi(\textbf{r}_\perp)
\end{equation}
Note that since we have already performed the Lorentz transformation of the fields, under the variable substitution $\tilde{z} = \gamma (z + \beta c t)$ we can write
\begin{equation}
   \mathcal{G}(\tilde{z},\tilde{\sigma}_z) = \frac{1}{\gamma}\  \mathcal{G}(z + \beta c t,\sigma_z)
\end{equation}
where $\mathcal{G}(z,\sigma_z)$ is now the longitudinal charge density as seen in the lab frame O. Further, since $\tilde{\textbf{r}}_\perp = \textbf{r}_\perp$, the force seen by a particle going through the potential $\Phi$ in the laboratory frame is thus
\begin{equation}  
  \textbf{F}_\perp = - \frac{N e^2 (1+\beta^2)}{4 \pi \varepsilon_0} \mathcal{G}(z+\beta c t,\sigma_z)\boldsymbol{\nabla}_\perp \Psi
\end{equation}

Taking $\beta \simeq 1$ for ultrarelativistic beams, we convert from spatial ($z$) to temporal ($t$) coordinate using
\begin{equation}
    \mathcal{G}(z+ct,\sigma_z) = \frac{1}{2c}\ \mathcal{G} \bigg(\frac{t}{2}+\frac{z}{2c},\sigma_t \bigg)
\end{equation}
where $\sigma_t = \sigma_z / (2c)$. Extending this result to a sum over multiple such potentials spaced in $z$ at intervals of $j \lambda$, and making the substitution of Eq. (\ref{eqn:potential}), we have
\begin{equation}
      \textbf{F}_\perp = - \frac{2 N e^2}{4 \pi \varepsilon_0 c} \sum_{j=1}^M \mathcal{G} \bigg(\frac{t}{2}+\frac{z-(j-1)\lambda}{2c},\sigma_t \bigg)\frac{\textbf{r}_\perp}{\sigma_j^2}
\end{equation}
With the parameter definition of $\Omega^2$ in Eq. (\ref{eqn:omega}) we see that this is equivalent to Eq. (\ref{eqn:Fperp}).

\section{RRF Equations of Motion}
\label{appendixD}

Starting from the relativistic Lorentz-Abraham-Dirac (LAD) equation of motion given by Eq. (\ref{eqn:LAD}-\ref{eqn:Frad}), we recognize that the right-most term in the expression for $\mathcal{F}_\mu^\text{rad}$ is the Lorentz-invariant form of the radiated Larmor power
\begin{equation}
    \mathcal{P} = - \tau_e \left[\frac{dp_\mu}{d\tau} \frac{dp^\nu}{d\tau} \right]
\end{equation}
where $p_\mu = (U/c, -\textbf{p})$, $U = \gamma m c^2$ is the particle energy, $\tau_e \equiv 2 r_e / (3 c)$, and $\textbf{p}$ is the usual 3-vector momentum. The external  4-vector force is given in component form by
\begin{equation}
    \mathcal{F}_\mu^\text{ext} = \gamma (\pmb{\beta} \cdot \textbf{F}, -  \textbf{F})
\end{equation}
where $\textbf{F} = -e (\textbf{E} + \pmb{\beta} \times \textbf{B})$ is the usual 3-vector Lorentz force in the lab frame and $\pmb{\beta} = \textbf{v}/c$ is the normalized velocity. In component form, the full LAD equations (\ref{eqn:LAD}) then take the following form, with no approximations:
\begin{equation}
\begin{aligned}
    \dot{U} &= [\tau_e (\gamma \ddot{U} + \dot{\gamma} \dot{U}) - \mathcal{P}] + \textbf{v} \cdot \textbf{F} \\
    \dot{\textbf{p}} &= [\tau_e (\gamma \ddot{\textbf{p}}+\dot{\gamma} \dot{\textbf{p}})-\frac{\pmb{\beta}}{c} \mathcal{P}]+\textbf{F}
    \end{aligned}
    \label{eqn:LADcomponent}
\end{equation}
where we have employed $d\tau = dt/\gamma$ and overdots denote derivatives with respect to $t$. The Larmor power is given by
\begin{equation}
    \mathcal{P} = \gamma^2 \tau_e [\dot{\textbf{p}}^2 - \dot{U}^2/c^2]
\end{equation}

Let us introduce the following normalized variables:
\begin{equation}
    q^\nu \equiv \frac{p^\nu}{m c}=(\gamma, \textbf{q}) \ \;\ \ \textbf{q} \equiv \gamma \pmb{\beta} \ \ ;\ \ \bar{\mathcal{P}} \equiv \frac{\lambda}{c} \frac{\mathcal{P}}{m c^2}
\end{equation}
\begin{equation}
    \bar{\textbf{F}}\equiv \frac{\lambda}{c} \frac{\textbf{F}}{m c} \ \  ; \ \ \alpha_e \equiv \frac{c \tau_e}{\lambda}
\end{equation}
Switching to derivatives with respect to the normalized time $u = c t/\lambda$, which we will denote by primes instead of dots, we thus have in place of Eq. (\ref{eqn:LADcomponent}),
\begin{equation}
    \begin{aligned}
    \gamma' &= [\alpha_e (\gamma \gamma''+\gamma'^2)-\bar{\mathcal{P}}]+\pmb{\beta} \cdot \bar{\textbf{F}} \\
    \textbf{q}' &= [\alpha_e (\gamma \textbf{q}'' + \gamma' \textbf{q}') - \pmb{\beta} \bar{\mathcal{P}}]+\bar{\textbf{F}}
    \end{aligned}
    \label{eqn:LADprimes}
\end{equation}
with
\begin{equation}
    \bar{\mathcal{P}} = \gamma^2 \alpha_e [\textbf{q}'^2 - \gamma'^2]
    \label{eqn:Pbar}
\end{equation}
Thus far no approximations have been made and these equations are quite general. We note that the left-most expressions within Eq. (\ref{eqn:LADprimes}) are small corrections. Even in the extreme case where the particle loses all of its energy, neglecting these terms produces only about a 10\% error. We thus omit them and combine Eqs. (\ref{eqn:LADprimes}) and (\ref{eqn:Pbar}), employing the approximation of Landau and Lifshitz, whereby we assume to lowest order that on the right side we may substitute the relativistic forms for the momentum and energy derivatives with the expressions in the absence of RRF ($\textbf{q}' = \bar{\textbf{F}}$, $\gamma' = \pmb{\beta} \cdot \bar{\textbf{F}}$), yielding

\begin{equation}
    \begin{aligned}
    \gamma' &\simeq \pmb{\beta} \cdot \bar{\textbf{F}}-\alpha_e \gamma^2 [\bar{\textbf{F}}^2-(\pmb{\beta} \cdot \bar{\textbf{F}})^2] \\
    \textbf{q}' &\simeq \bar{\textbf{F}}-\alpha_e \pmb{\beta} \gamma^2 [\bar{\textbf{F}}^2 - (\pmb{\beta} \cdot \bar{\textbf{F}})^2]
    \label{eqn:LADsimp2}
    \end{aligned}
\end{equation}

Taking the force to be predominantly transverse, in accordance with our earlier prescription ($\textbf{F} = \textbf{F}_\perp$) and noting that in the limit of small transverse velocity $\beta_\perp \ll \beta_z \simeq 1$, to good approximation,
\begin{equation}
    \bar{\textbf{F}}_\perp^2 - (\pmb{\beta} \cdot \bar{\textbf{F}}_\perp)^2 \simeq \bar{\textbf{F}}_\perp^2 \beta^2
    \label{eqn:Fapprox}
\end{equation}
Upon insertion of (\ref{eqn:Fapprox}) into Eqs. (\ref{eqn:LADsimp2}), we immediately obtain Eqs. (\ref{eqn:RRFeqnmotion}).

\section{Envelope Equation with RRF}
\label{appendixE}
The RRF equations of motion Eq. (\ref{eqn:RRFeqnmotion}) cast into unnormalized variables read as follows:
\begin{equation}
    \begin{aligned}
    \dot{\gamma} =& \ \frac{\textbf{v} \cdot \textbf{F}_\perp}{m c^2} - \tau_e \frac{p^2}{(m c)^2} \left( \frac{\textbf{F}_\perp}{m c} \right)^2 \\
    \dot{\textbf{p}} =& \ \textbf{F}_\perp - \tau_e \gamma \textbf{p} \left(\frac{\textbf{F}_\perp}{m c}\right)^2
    \end{aligned}
\end{equation}
where $\textbf{p} = \gamma m \textbf{v}$ is the usual 3-vector momentum. Taking the external force to have the previously derived form $\textbf{F}_\perp = - \gamma_0 m K^2 \textbf{r}_\perp$ where $\textbf{r}_\perp = x \hat{\textbf{x}} + y \hat{\textbf{y}}$, the above equations take the component forms
\begin{equation}
    \begin{aligned}
    \ddot{x} &= - \frac{\gamma_0}{\gamma} K^2 x - \frac{\dot{\gamma}}{\gamma} \dot{x} - \dot{x} \frac{\tau_e \gamma \gamma_0^2 K^4 (x^2+y^2)}{c^2} \\
    \ddot{y} &= - \frac{\gamma_0}{\gamma} K^2 y - \frac{\dot{\gamma}}{\gamma} \dot{y} - \dot{y} \frac{\tau_e \gamma \gamma_0^2 K^4 (x^2+y^2)}{c^2} \\
    \dot{\gamma} &= -\gamma_0 \frac{K^2}{c^2} (\dot{x}x+\dot{y}y) - \tau_e \gamma^2 \beta^2 \left[\frac{\gamma_0^2 K^4 (x^2+y^2)}{c^2}\right]
    \end{aligned}
    \label{eqn:xygamma}
\end{equation}
We now switch to cylindrical coordinates, taking $x = \sigma \cos \phi$, $y = \sigma \sin \phi$. We here take the particle to correspond to the RMS particle at radial position $\textbf{r}_\perp = \sigma \hat{\textbf{r}}$. We ignore the factor of $\beta^2$ in the equation for $\gamma$ as it has a negligible effect. Multiplying the $x$ equation by $\cos \phi$ and the $y$ equation by $\sin \phi$ and adding them, we obtain
\begin{equation}
    \begin{aligned}
    &\ddot{\sigma}+\left( \frac{\gamma_0}{\gamma} K^2 - \dot{\phi}^2 \right) \sigma + \frac{\dot{\gamma}}{\gamma} \dot{\sigma} + \gamma \left[\frac{\tau_e \gamma_0^2 K^4}{c^2} \right]\dot{\sigma} \sigma^2 = 0 \\
    &\dot{\gamma} + \gamma_0 \frac{K^2}{c^2} \dot{\sigma} \sigma + \frac{\tau_e \gamma_0^2 K^4}{c^2} \sigma^2 \gamma^2 = 0
    \end{aligned}
    \label{eqn:sigmagamma}
\end{equation}
We additionally note that the angular momentum $\textbf{L} = \textbf{r} \times \textbf{p}$ has longitudinal component $L_z = x p_y - y p_x = \gamma m \sigma^2 \dot{\phi}$. Taking the time derivative, we may obtain a third equation of motion:
\begin{equation}
    \dot{L}_z = \gamma m x \left[\frac{\dot{\gamma}}{\gamma}\dot{y}+\ddot{y}\right]-\gamma y \left[\frac{\dot{\gamma}}{\gamma} \dot{x} + \ddot{x}\right]
\end{equation}
Inserting into this expression from the equations of motion for $x$ and $y$ in Eqs. (\ref{eqn:xygamma}), we find that this equation is equivalent to
\begin{equation}
    \dot{L}_z = - \left[\frac{\tau_e \gamma \gamma_0^2 K^4 \sigma^2}{c^2}\right] L_z
    \label{eqn:Lz}
\end{equation}
Thus, in the absence of RRF, this becomes $\dot{L}_z = 0$ and hence the longitudinal angular momentum is a constant of the motion. We associate this constant with the normalized transverse emittance, which for linear focusing and acceleration in the absence of RRF is a conserved quantity: $L_z/(m c) = \gamma \epsilon = \epsilon_n$. In accordance with Eq. (\ref{eqn:Lz}) this invariance can no longer be assumed, and in fact $\epsilon_n$ can be seen to \textit{decrease} during the interaction. Converting Eqs. (\ref{eqn:sigmagamma},\ref{eqn:Lz}) to normalized derivatives with respect to $u = (\lambda/c) t$ and invoking the normalized variables,
\begin{equation}
\begin{aligned}
    \bar{\sigma} \equiv \frac{\sigma}{\sigma_0} \ \ ; \ \ \bar{\gamma} \equiv & \frac{\gamma}{\gamma_0} \ \ ; \ \ \hat{K}= \frac{\lambda}{c} K \\
    \Lambda \equiv \frac{\lambda L_z}{\gamma_0 m c \sigma_0^2} = \frac{\lambda \epsilon_n}{\gamma_0 \sigma_0^2} \ \ &; \ \ \bar{\alpha}_e \equiv \frac{c \tau_e}{\lambda} \gamma_0^3 \left( \frac{\sigma_0}{\lambda} \right)^2
    \end{aligned}
\end{equation}
we may recast Eqs. (\ref{eqn:sigmagamma},\ref{eqn:Lz}) into the form
\begin{equation}
    \begin{aligned}
    &\bar{\sigma}'' + \frac{1}{\bar{\gamma}}\hat{K}^2 \bar{\sigma}-\frac{\Lambda^2}{\bar{\gamma}^2 \bar{\sigma}^3}+\frac{\bar{\gamma}'}{\bar{\gamma}} \bar{\sigma}' + \bar{\alpha}_e \hat{K}^4 \bar{\sigma}' \bar{\sigma}^2 \bar{\gamma} = 0 \\
   &\bar{\gamma}' + \left( \frac{\sigma_0}{\lambda} \right)^2 \bar{\sigma}' \bar{\sigma} \hat{K}^2 + \bar{\alpha}_e \hat{K}^4 \bar{\sigma}^2 \bar{\gamma}^2 = 0 \\
   &\Lambda' + \bar{\alpha}_e [\bar{\gamma} \hat{K}^4 \bar{\sigma}^2] \Lambda = 0
    \end{aligned}
    \label{eqn:LADenvfull}
\end{equation}
subject to the initial conditions
\begin{equation}
    \bar{\sigma}(0) = 1 \ ; \ \bar{\sigma}'(0) = 0 \ ; \ \bar{\gamma}(0) = 1 \ ; \ \Lambda(0) = \Theta
    \label{eqn:initcond}
\end{equation}
Here $\Theta$ is the normalized emittance term defined in Eq. (\ref{eqn:defs_theta}). We observe that the second term containing $(\sigma_0 / \lambda)^2$ in the equation for $\bar{\gamma}$ in Eq. (\ref{eqn:LADenvfull}) is small and has negligible effect on the solution. (Neglecting it produces errors on the order of 1 part in $10^6$). Hence, from the second and third relations of (\ref{eqn:LADenvfull}), we find that 
\begin{equation}
\label{eqn:gammalambda}
\begin{aligned}
   &\bar{\gamma}' + [\bar{\alpha}_e \hat{K}^4 \bar{\sigma}^2] \bar{\gamma}^2 = 0 \\
   &\Lambda' + [\bar{\alpha}_e \hat{K}^4 \bar{\sigma}^2] \Lambda \bar{\gamma} = 0
    \end{aligned}
\end{equation}
Subject to the initial conditions (\ref{eqn:initcond}), these yield the solution $\Lambda(u) = \Theta \bar{\gamma}(u)$. From the definitions of the parameters, we see that this is equivalent to $\epsilon_n(u) = \epsilon_0 \gamma(u)$, where $\epsilon_0$ is the initial RMS emittance. Hence we obtain the remarkable result that the normalized emittance is not conserved under the action of the RRF but instead varies directly with $\gamma$. This observation negates the need for the third equation in (\ref{eqn:LADenvfull}). Upon making the substitution $\Lambda = \Theta \bar{\gamma}$ into the first of Eq. (\ref{eqn:LADenvfull}), neglecting the second term in the equation for $\bar{\gamma}$ per the discussion above, and restoring the subscripted $i$'s to denote that the equations are to be solved separately for each microbunch in the train, we immediately obtain Eqs. (\ref{eqn:RRFenvelope}). 

\section{Energy Correction Factor}
\label{appendixC}

We note that the energy equation of motion (\ref{eqn:RRFenvelope}) with the $\pmb{\beta} \cdot \bar{\textbf{F}}$ terms neglected amounts to equating the energy derivative with the Larmor radiation:
\begin{equation}
   \gamma' = - \bar{\mathcal{P}} \ \ ;  \ \ \text{where}\ \  \bar{\mathcal{P}} = \gamma^2 \alpha_e \bar{\textbf{F}}_\perp^2 \ .
\end{equation}

If we consider the full form of the potential $\Psi$ from Eq. (\ref{eqn:potential}), the transverse force for the $(i,j)$ bunch crossing is
\begin{equation}
    \bar{\textbf{F}}_\perp = -\frac{1}{2}(\gamma_0/\lambda) \hat{K}_{ij}^2 \sigma_j^2 \nabla_\perp \Psi_j
\end{equation}
where $\nabla_\perp \Psi_j = \nabla_\perp \Psi |_{\sigma \rightarrow \sigma_j}$ from Eq. (\ref{eqn:potential}). Upon insertion into the energy equation we obtain the energy loss of interaction $(i,j)$
\begin{equation}
    \bar{\gamma}_{ij}' = - \bar{\alpha}_e \bar{\gamma}_i^2 \bar{\sigma}_i^2 \hat{K}_{ij}^4 \times \frac{1}{4} \frac{\sigma_j^2}{\sigma_i^2} \left| \sigma_j \boldsymbol{\nabla}_\perp \Psi_j \right|^2
\label{eqn:gammaprimecorrected}
\end{equation}
We recognize this is the same as our form of the energy equation of motion in Eq. (\ref{eqn:RRFenvelope}) but with an extra factor on the right. Integrating over the normalized time $u$ and the transverse distribution $F_i(\textbf{r}_\perp)$ of the $i$'th incident bunch, the average energy loss for the $(i,j)$ interaction is
\begin{equation}
   \delta \bar{\gamma}_{ij} = - \int \bar{\gamma}_{ij}' F_i(\textbf{r}_\perp) d^2\textbf{r}_\perp du 
   \end{equation}
With the form of Eq. (\ref{eqn:gammaprimecorrected}), we see that this will yield the same result as Eq. (\ref{eqn:dgammaint}) but with an extra dimensionless factor
\begin{equation}
    \mathcal{Z}_{ij} \equiv \frac{1}{2} \int [\mathcal{G}(r,\sigma_j)]^2 \frac{\sigma_j^2}{\sigma_i^2} \left| \sigma_j \boldsymbol{\nabla}_\perp \Psi_j \right|^2 d^2\textbf{r}_\perp
\label{eqn:ZijDef}
\end{equation}
Making the variable substitutions $\zeta = \sigma_j r$, $\mu = \sigma_j / \sigma_i$ and taking $d^2\textbf{r}_\perp = r dr d\phi \rightarrow 2 \pi \sigma_j^2 \zeta d\zeta$ this integral becomes
\begin{equation}
\label{eqn:Zijintegral}
    \mathcal{Z}_{ij} = 2 \mu^4 \int_0^\infty e^{-\mu^2 \zeta^2} \frac{1}{\zeta} \left[1-e^{-\zeta^2}\right]^2 d\zeta
\end{equation}    
which evaluates explicitly to 
\begin{equation}
\label{eqn:Factor}
    \mathcal{Z}_{ij} = 2 \mu^4 \ln \left[\frac{1+\mu^2}{\mu \sqrt{2+\mu^2}} \right] \ \ ; \ \ \mu \equiv \frac{\sigma_j}{\sigma_i}
\end{equation}   
To reconcile this with our prior result, Eq. (\ref{eqn:dgammaij}), in the absence of the extra factor of $\mathcal{Z}_{ij}$, we note that the linear force approximation amounts to taking the term in square brackets in Eq. (\ref{eqn:Zijintegral}) $[...] \approx \zeta^2$ to lowest order and the integral in this approximation evaluates to $\mathcal{Z}_{ij} = 1$. We further note that in the case of equal bunch sizes ($\mu = 1$), $\mathcal{Z}_{ij} = 2 \ln (2/\sqrt{3}) \simeq 25 \pi/288$. For convenience of other calculations, we may thus write the correction factor in the form
\begin{equation}
    \mathcal{Z}_{ij} \simeq \frac{25 \pi}{288} \bar{Z}_{ij} \ ; \ \text{where} \ \bar{Z}_{ij} \equiv \frac{\mathcal{Z}_{ij}}{2 \ln (2/\sqrt{3})}
    \label{eqn:Zijnorm}
\end{equation}


\bibliography{england}

\end{document}